\documentclass[12pt]{article}

\usepackage{graphicx}
\usepackage{epsfig}
\usepackage{amsmath}
\usepackage{mathrsfs}
\usepackage{amssymb}
\usepackage{amsthm}
\usepackage[super]{natbib}
\usepackage{algorithm,algorithmic}
\usepackage{url}
\usepackage[margin=2.5cm]{geometry}
\usepackage{rotating}
\usepackage{setspace}
\usepackage{stackrel}
\usepackage{authblk}

\usepackage{xr}
\makeatletter
\newcommand\code{\bgroup\@makeother\_\@makeother\~\@makeother\$\@makeother\^\@codex}
\def\@codex#1{{\normalfont\ttfamily\hyphenchar\font=-1 #1}\egroup}

\newcommand\proglang{\bgroup\@makeother\_\@makeother\~\@makeother\$\@makeother\^\@codex}
\def\@codex#1{{\normalfont\ttfamily\hyphenchar\font=-1 #1}\egroup}

\newcommand\pkg{\bgroup\@makeother\_\@makeother\~\@makeother\$\@makeother\^\@codex}
\def\@codex#1{{\normalfont\ttfamily\hyphenchar\font=-1 #1}\egroup}
\makeatother

\newcommand{\csp}[1]{\mathcal{#1}}

\renewcommand{\P}{\mathbb{P}}

\newcommand{\E}{\mathbb{E}}

\newcommand{\cov}{\mathrm{cov}}

\newcommand{\prob}{\mathbb{P}}

\begin{document}

%\title{GPU-Accelerated Real-Time Downscaling of Malaria Incidence in Zambia}
\title{Malaria Risk Mapping Using Routine Health System Incidence Data in Zambia}
% Ricardo Andrade-Pacheco, Hugh Sturrock, Adam Bennett
% MEI, Global Health Group, UCSF, San Francisco, CA, US
%
% Busiku Hamainza
% Zambia National Malaria Control Centre, Lusaka, Zambia
%
% Kafula Silumbe, John Miller
% Malaria Control and Elimination Partnership in Africa, Lusaka, Zambia
%
% Thomas P. Eisele
% Tulane University, New Orleans, LA, US
\author[1]{\small Benjamin M. Taylor} % benjamin.taylor3@nhs.net
\author[2]{\small Ricardo Andrade-Pacheco} %ric70x7@gmail.com
\author[2]{\small Hugh Sturrock} % Hugh.Sturrock@ucsf.edu
\author[3]{\small Busiku Hamainza} % bossbusk@gmail.com
\author[4]{\small Kafula Silumbe} % ksilumbe@path.org
\author[4]{\small John Miller} % jmiller@path.org
\author[5]{\small Thomas P. Eisele} % teisele@tulane.edu
\author[2]{\small Fran\c{c}ois Rerolle} % Francois.Rerolle@ucsf.edu
\author[6]{\small Hannah Slater} % hslater@path.org
\author[2]{\small Adam Bennett} % adam.bennett@ucsf.edu

\affil[1]{\footnotesize (Corresponding Author) Blackpool Teaching Hospitals NHS Foundation Trust, correspondence to: benjamin.taylor3\@nhs.net}
\affil[2]{\footnotesize Malaria Elimination Initiative, Institute for Global Health Sciences, UCSF, San Francisco, CA, US}
%\affil[3]{\footnotesize Imperial College London, London, UK}
%\affil[4]{\footnotesize Big Data Institute, Nuffield Department of Medicine, University of Oxford, Oxford, UK}
\affil[3]{\footnotesize Zambia National Malaria Control Centre, Lusaka, Zambia}
\affil[4]{\footnotesize Malaria Control and Elimination Partnership in Africa, Lusaka, Zambia}
\affil[5]{\footnotesize Tulane University, New Orleans, LA, US}
\affil[6]{\footnotesize Imperial College, London, UK}
\maketitle

% B.T., R.A.-P., H.S., A.B. wrote the manuscript
% B.T. implemented the algorithms and conducted the incidence modelling
% R.A-P. H.S. A.B. F.R. did the catchment modelling
% B.H., K.S., J.M., T.E. assisted with data capture
% All Authors Reviewed the Manuscript

\begin{abstract}
     Improvements to Zambia's malaria surveillance system allow better monitoring of incidence and targetting of responses at refined spatial scales. As transmission decreases, understanding heterogeneity in risk at fine spatial scales becomes increasingly important. However, there are challenges in using health system data for high-resolution risk mapping: health facilities have undefined and overlapping catchment areas, and report on an inconsistent basis. We propose a novel inferential framework for risk mapping of malaria incidence data based on formal down-scaling of confirmed case data reported through the health system in Zambia. We combine data from large community intervention trials in 2011-2016 and model health facility catchments based upon treatment-seeking behaviours; our model for monthly incidence is an aggregated log-Gaussian Cox process, which allows us to predict incidence at fine scale. We predicted monthly malaria incidence at 5km$^2$ resolution nationally: whereas 4.8 million malaria cases were reported through the health system in 2016, we estimated that the number of cases occurring at the community level was closer to 10 million. As Zambia continues to scale up community-based reporting of malaria incidence, these outputs provide realistic estimates of community-level malaria burden as well as high resolution risk maps for targeting interventions at the sub-catchment level.
\end{abstract}

\section{Background}

Improving understanding of the spatial distribution of malaria incidence and prevalence is critical for the design and implementation of effective monitoring, control and elimination strategies. Particularly with decreasing transmission, there is greater need to measure and visualize heterogeneity in risk and target interventions at the community level. The wide availability of household surveys measuring parasite prevalence, coupled with advances in geostatistical modeling and computational techniques has allowed the development of high-resolution risk maps. However, these surveys are infrequent, often prohibitively expensive, and inefficient for measuring disease burden, especially in lower transmission settings.

Routine confirmed case data collected through the health systems are an under-utilized data source for risk mapping. These data present a tremendous opportunity for high resolution risk mapping given their wide geographic and temporal coverage and relative inexpensive collection. Although health management information systems are rapidly improving in sub-Saharan Africa, few malaria programs are able to routinely geo-locate cases at levels below the reporting facility, and as a result case counts are aggregated at the facility levels. There are many biases associated with using facility locations and aggregate case counts and associated covariates to develop high-resolution maps of incidence or risk, including differential rates of treatment seeking, unknown catchment boundaries, and inconsistent reporting. For understanding and mapping incidence in space and time, facility-level data often remain the only option, and the issues related to these aggregated data have been dealt with in various, often suboptimal, ways.

Zambia has achieved dramatic success in reducing its malaria burden through scale up of vector control, community case management, and surveillance systems, and is ambitiously targeting sub-national elimination. Zambia has enabled a more spatially refined understanding of incidence through the scale-up of reporting through community health workers (CHW) in parts of the country, but this scale-up has been staggered and large areas of the country have not yet been reached. Additionally, CHW reporting data are subject to the same biases as health facility data: they are often incomplete, depend upon treatment-seeking behavior, and are themselves still spatially aggregated - albeit at smaller scales. As a result of these limitations, generating fine-scale incidence-based risk maps that enable identification of transmission foci has to date not been thoroughly examined, although recent work on down-scaling has been encouraging \citep{sturrock2014, alegana2012spatial, alegana2016advances, weiss2019mapping}. With improved health management information system (HMIS) data reported through DHIS2 (District Health Information System 2) and now several years of scale-up in the CHW reporting system, combined with novel spatial statistical approaches \citep{taylor2013,taylor2015, taylor2018} there is a unique opportunity to revisit the challenge of modelling risk using aggregated health system data.

A potential modelling approach for count data pertaining to discrete spatial units, such as health facilities or community healthcare workers, involves using an aggregated log-Gaussian Cox process (ALGCP), see \cite{taylor2018}. This model assumes a spatially continuous latent log-Gaussian Cox process (LGCP) for the (unobserved) true locations of cases and incorporates aggregation mechanisms through a data augmentation step. However this approach has not yet addressed variable health facility choice and overlapping catchment boundaries that change over time, primarily due to issue of the computational cost of such large scale mapping exercises.

In this article, we extend the modeling approach described in \cite{taylor2018} by: (i) developing a model of health facility choice from household-level treatment-seeking data; (ii) developing an algorithm for inference with space-time data with fuzzy and overlapping catchment boundaries; (iii) detailing how to correctly perform inference in the presence of missing data; and (iv) providing an open-source highly-parallel NVIDIA graphics processing unit (GPU) implementation of our algorithm that is an order of magnitude faster compared with a single-core version. We apply these methods to a country-wide analysis of monthly malaria incidence in Zambia between 2012 and 2016, accounting for differential treatment-seeking rates, inconsistent reporting, and overlapping and poorly-defined catchment areas. The outputs will be used to complement the improved health system data in Zambia and provide information for targeting intervention strategies at the community level.

\section{Methods}

\subsection{Data Description}

\subsubsection{Treatment-seeking data}

\emph{MTAT intervention data}

Treatment-seeking and health facility choice data were incorporated from a community randomized controlled trial evaluating mass test and treat conducted in 2011-2013 in Southern Provence \cite{larsen2015population} where all individuals in intervention areas were offered an RDT test by their home and treated with ACT if testing positive. During this trial over 720,600 individual observations were included, of which 60,164 reported fever in the previous two weeks, and 16,649 reported seeking treatment for fever. Individuals who sought treatment were also asked which facility they sought care from; individuals sought care from a total of 50 different facilities in Southern Province, including 49 health centers and 1 health post.

\emph{Household survey data}

Four national malaria indicator surveys (MIS) (2010, 2011, 2012 and 2013) and three MIS surveys in Southern Province (2014, 2015, 2016) were conducted over the time period of study. During these surveys, all individuals were asked about treatment seeking for fever in the previous two weeks, and if they sought care, what type of facility they sought care from (public, private, CHW, etc.). For the national MIS surveys, the total number of individuals who sought treatment out of those who had fever was: 843 out of 1297 in 2010; 849 out of 1284 in 2011; 517 out of 829 in 2012; and 2099 out of 2743 in 2013. For the Southern Province surveys, the total number of individuals who sought treatment out of those who had fever was: 1417 out of 2315 in 2014; 585 out of 1030 in 2015; and 547 out of 976 in 2016.

\subsubsection{Confirmed malaria case data}

\emph{HMIS data}

Confirmed malaria case data are reported on a monthly basis into DHIS2 at all health facilities. There were a total of 2119 reporting facilities between 2012 and 2016. GPS data were available for 1,334 (63\%). Facilities without a GPS point were removed from the analysis.

\emph{CHW reporting data}

Community case management through CHWs was implemented in 2013 in Southern Province and scaled up rapidly over the study period. CHWs report data on the number of RDT tests conducted and number of positive cases per month into DHIS2. Following initial implementation, the number of reporting CHWs increased over the study period from 1457 in 2013 to 2,239 in 2016.

\subsubsection{Covariate data}

\emph{Incidence (risk) model}

Table \ref{tab:covariates} gives details of the covariates we used in the incidence model. With the exception of daytime temperature, we assumed the relationship between these covariates and case incidence was linear on the log-risk scale.

% [1] "Intercept"     "Rainfall"      "Elevation"     "EVI"
% [5] "Impervious"    "TempDay"       "TempNight"     "TWI"
% [9] "Waterbodies"   "LogPopulation"

\begin{table}[htbp]
    \footnotesize
    \begin{tabular}{|l|l|l|}
        Variable Name & Source & Detail\\ \hline
        Rainfall & http://chg.geog.ucsb.edu/data/chirps/ & Monthly precipitation (5km) \\
        % Elevation & SRTM & elevation above sea level (1km) \\
        EVI & MODIS & Monthly enhanced vegetation index (1km) \\
        Temp. Day & MODIS & av. monthly day time temperature (1km) \\
        Temp. Night & MODIS & av. monthly night time temperature (1km) \\
        TWI &  & topographic wetness index \\
        % Dist. to water bodies & DIVA-GIS & (Km) \\
    \end{tabular}
    \caption{\label{tab:covariates} Showing details of the covariate data used in our analyses.}
\end{table}

\begin{figure}[htbp]
    \centering
    \includegraphics[width=0.5\textwidth]{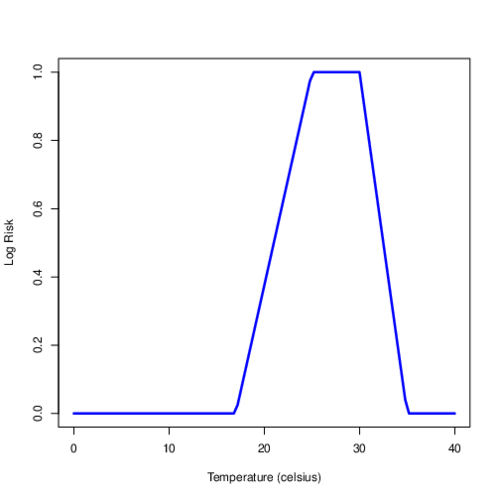}
    \caption{\label{fig:daytemp} Showing assumed relationship between risk on the log scale and daytime monthly average temperature.}
\end{figure}

Figure \ref{fig:daytemp} shows the assumed functional relationship between log risk and daytime temperature. Consistent with epidemiological knowledge,
we assumed that the optimal malaria transmission occurs between between 25--30 degrees Celsius and that there would be no transmission below 17 degrees or above 35
degrees;
we encoded this information as a piecewise linear function and applied this function to the daytime temperature data by pixel \citep{mordecai2013optimal, parham2009modeling, hoshen2005model, Paaijmans15135}. Missing data in day and night time temperatures were filled in using a random forest model with elevation as a covariate alongside $x$, $y$ cordinates and time (code available from \url{https://github.com/HughSt/gapfilling_rasters}).

\subsection{Treatment seeking model}

We assumed that the probability that an individual with fever seeks
treatment is primarily driven by the distance and cost of travelling to a
health facility (hospital, health center or health post).
We also assumed that individuals choose their route to a health facility
based on the obstacles and friction presented by the landscape,
as opposed to a straight line.

The travel cost of a given route between points was modeled as a function that weights distance and friction.
We used the R-package gdistance and followed the parametrization suggested by
\cite{etten2017gdistance} to define the transition costs of moving across
Zambia. To define the landscape, we used a set of geographic layers including
elevation (90m resolution from SRTM), land cover (300m from GlobalCover project), roads (primary, secondary and tertiary from Open Street Map) and major rivers. Once the transition costs were defined, the least-cost routes were
found with Djiskstra's algorithm using the R-package igraph
\cite{csardi2006igraph}. As we would expect, plots of the travel cost (not shown here) show changes over
time in travel cost result from new health facilities or CHWs being active.

With an estimate of the travel cost for each individual in the Southern Province
MTAT data, we defined a geostatistical model for their decision to seek treatment
as a spatio-temporal Bernoulli process with parameter $\theta(s,t)$, such that

\begin{equation*}
  \mbox{logit}(\theta(s,t)) = \beta_0 + \beta_1 r(s,t) + f_1(s)  + f_2(t),
\end{equation*}
where $r(s,t)$ is the travel cost of the least-cost route from a household with
location $s$ at time $t$; $f_1(s)$ is a spatial random process with variance $\sigma^2_s$ and nominal range $\ell_s$; and $f_2(t)$ is a random walk across time with variance $\sigma^2_t$. We fitted this model using the R package R-INLA. Table
\ref{tab:tseeking-param} shows the parameter estimates of this model.

\begin{table}[htbp]
    \footnotesize
    \centering

    \begin{tabular}{|l|l|l}
        Variable          & Mean    & 95\% Confidence Interval\\
        \hline
        Intercept          &  0.46 & (-0.20,  1.13) \\
        Travel Cost        & -0.30 & (-0.39, -0.22) \\
        $\sigma^2_t$       &  0.36 & ( 0.06,  1.25) \\
        $\sigma^2_s$       &  5.43 & ( 0.60,  14.63) \\
        $\ell_s$           &  0.11 & ( 0.03,  0.20) \\
    \end{tabular}
    \caption{\label{tab:tseeking-param} Parameter estimates of the treatment seeking model.}
\end{table}

\begin{figure}
    \begin{center}
        \begin{minipage}{0.4\textwidth}
            \includegraphics[width=0.9\textwidth]{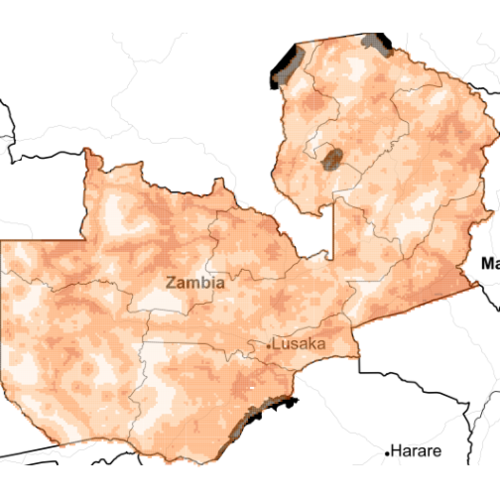}
        \end{minipage}\begin{minipage}{0.4\textwidth}
            \includegraphics[width=0.9\textwidth]{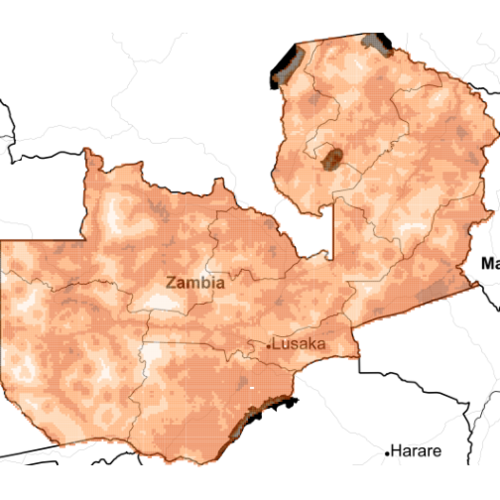}
        \end{minipage}

        \begin{minipage}{0.4\textwidth}
            \includegraphics[width=0.9\textwidth]{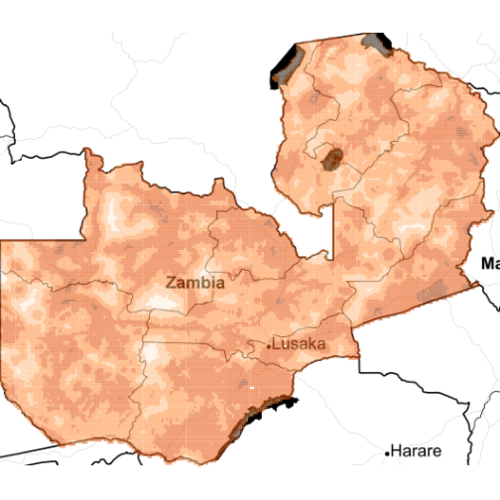}
        \end{minipage}\begin{minipage}{0.4\textwidth}
            \includegraphics[width=0.9\textwidth]{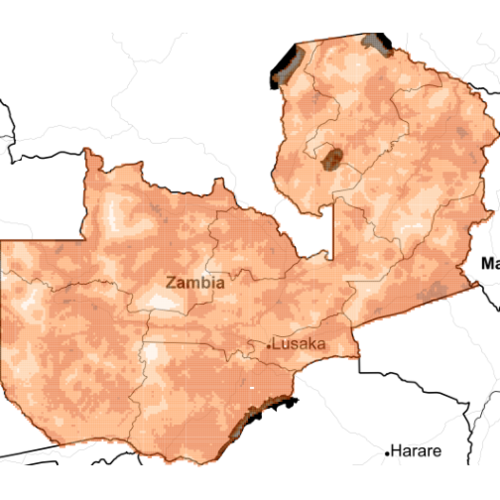}
        \end{minipage}

        \begin{minipage}{0.4\textwidth}
            \includegraphics[width=0.9\textwidth]{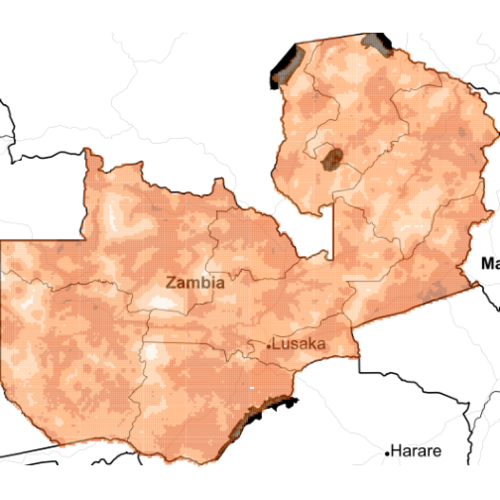}
        \end{minipage}

    \end{center}
    \caption{\label{fig:T_seeking} Probability of seeking treatment, based on the household location, left to right, top to bottom, the plots are for 2012-2016. Colours are:
    \protect\includegraphics[width=1em]{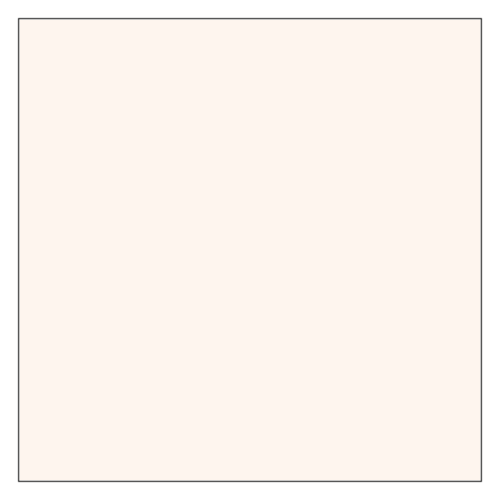} [ 0 , 0.2 ],
    \protect\includegraphics[width=1em]{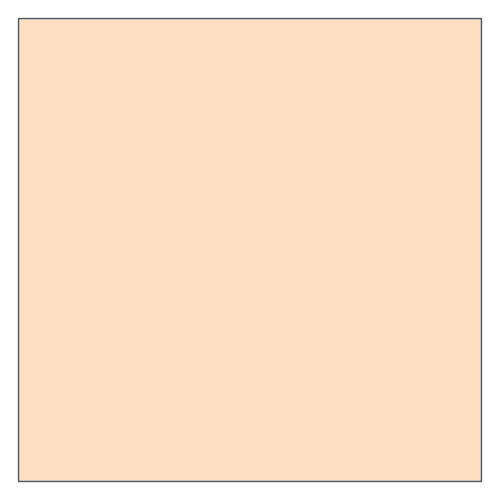} [ 0.2 , 0.4 ],
    \protect\includegraphics[width=1em]{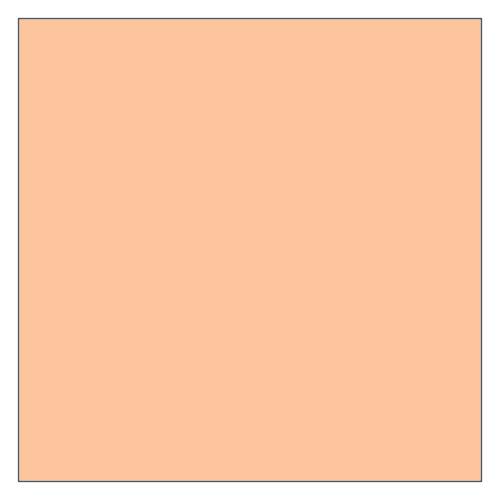} [ 0.4 , 0.6 ],
    \protect\includegraphics[width=1em]{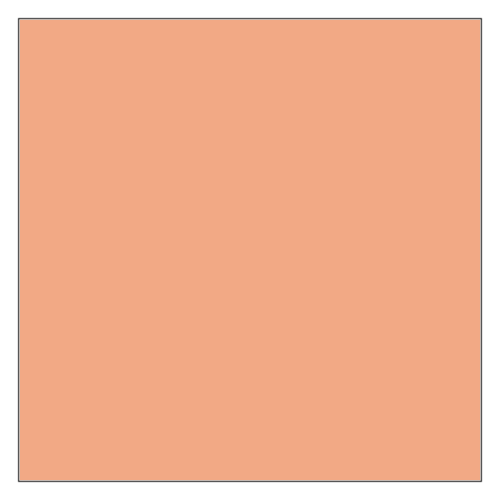} [ 0.6 , 0.8 ],
    \protect\includegraphics[width=1em]{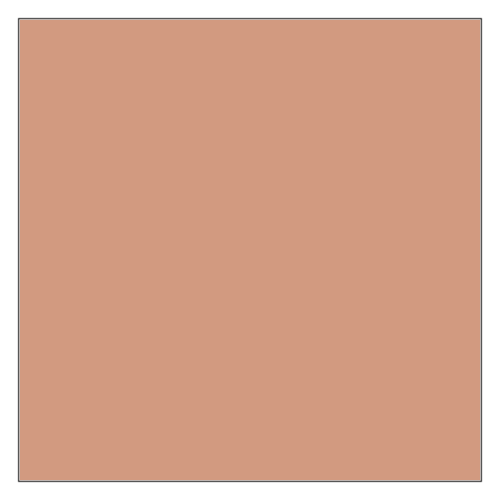} [ 0.8 , 1.0 ].}
\end{figure}

This model was generalized to define a surface of treatment seeking across the
country considering all the health facilities available. Figure
\ref{fig:T_seeking} shows probabilities of seeking treatment per year.

%A spatial plot of the treatment seeking probability is shown in Figure \ref{fig:WB_seek_covariates} (b).

\subsubsection{Health Facility Choice Model}

We assumed that the decision of which health facility to attend is based on a
comparison of their different attributes. Examples of such attributes are the
size of the health facility, the travel cost needed to get to it,
and previous experiences or recommendations from other patients.
To describe this selection
process we relied on the parameterization of Multiplicative Competitive Interaction (MCI) models, which are commonly used to describe spatial
interactions and share of market in retail \cite{cooper2010market,
huff2008calibrating}.

An individual at a particular location in space may have access to
a selection of health facilities. MCI models model each individual's utility, which
describes the attraction towards one of these possible choices, which can be defined as probabilities.

\begin{figure}[htbp]
    \centering
    \includegraphics[width=0.5\textwidth]{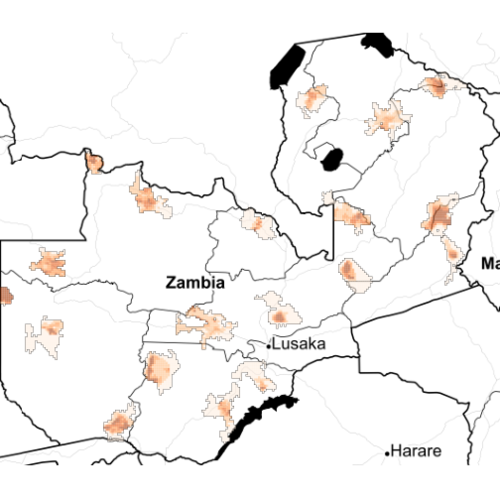}
    \caption{\label{fig:catchment_areas} Plot showing the cell-wise conditional probability of reporting (assuming one has chosen to seek treatment) to 19 randomly chosen health facility catchment areas spread out over the country (i.e. the $w_{ij}$ in Equation \ref{eqn:condprob}). Probabilities have been rescaled to $[0,1]$ for each facility i.e. for the $j$th facility, this is $w_{ij} / max_i\{w_{ij}\}$. Colours are:
    \protect\includegraphics[width=1em]{col1} [ 0 , 0.2 ],
    \protect\includegraphics[width=1em]{col2} [ 0.2 , 0.4 ],
    \protect\includegraphics[width=1em]{col3} [ 0.4 , 0.6 ],
    \protect\includegraphics[width=1em]{col4} [ 0.6 , 0.8 ],
    \protect\includegraphics[width=1em]{col5} [ 0.8 , 1.0 ].}
\end{figure}

We incorporate these probabilities into our modelling
framework by including an $n\times m$ weight matrix, $W$, for which the $[i,j]$
element represents the probability that an individual in cell $i$ will report to
facility $j$: the row sums of this matrix are therefore either 1 or 0 (in the
case that there are no individuals in cell $i$). The weight matrix can be
included as part of a multinomial step, which in addition needs to be adjusted
for the fact that facility catchment areas overlap, details in
\cite{taylor2018}. The union of grid cells whose indices correspond to non-zero
entries in the $j$th column of $W$ is the catchment area of health facility $j$,
as per Figure \ref{fig:catchment_areas}.

\subsubsection{Including Community Health Workers}

Inclusion of the community health care workers into our modelling framework was achieved by treating each CHW as if they were a small health facility. We  created a maximum potential catchment area for each CHW by using the grid cell the CHW was located in plus all cells in a neighbourhood of order 4 around that cell. We varied the probability of attendance at a CHW with the distance to the nearest health facility, such that attendance at a CHW was more likely if the nearest health facility was further away (see Appendix \ref{sect:huff_data}).

\subsection{Space-Time Aggregated Point Process Model}

To model incidence at the pixel level across Zambia, we assumed that the true locations of the cases follow a spatiotemporal log-Gaussian Cox process and following \cite{taylor2018}, that inference takes place on a fine grid. This model implies conditionally that the number of cases, $X(s,t)$, in a cell containing spatial location $s$ during the time window
containing $t$ is Poisson distributed:
\begin{eqnarray}
    X(s,t) &\sim& \text{Poisson}[R(s,t)]\\ \nonumber
    \log R(s,t) &=& \log P(s,t) + Z(s,t)\beta + \csp{Y}(s,t),
\end{eqnarray}
where $P(s,t)$ is a known component of the intensity function (population
density in our case), $Z(s,t)$ is a vector of covariates, $\beta$ is a
vector of parameter effects to be estimated and $\csp{Y}$ is a spatiotemporal
Gaussian process. We assume that $\cov[\csp{Y}(s_1,t_1),\csp{Y}(s_2,t_2)] =
\sigma^2\exp\{-||s_1-s_2||/\phi - |t_1-t_2|/\theta\}$ and that
$\E[\csp{Y}(s,t)] = -\sigma^2/2$ for all $s$ and $t$.

\subsubsection{Missing Values}

Our data contains a large number of missing values which correspond to active health facilities that have not reported case numbers for a given month. We adjusted our model to account for these missing values in a manner that respects the underlying model for the continuous process. We assumed that individuals who would normally use each of the facilities would continue to do so, even if the facility did not report in a particular month. Full details on model adjustments are available in Appendix \ref{sect:miss_val}.

\subsection{GPU Implementation}

Single or multiple CPU core implementations of the MCMC algorithm detailed above would be computationally very expensive, even using specialist linear algebra libraries. We therefore developed Graphics Processing Unit (GPU) code as an extension to the R package lgcp \citep{taylor2015} using the NVIDIA CUDA programming language and libraries for computational `grunt' and the R programming language for data handling and pre-processing. Computation for the main analysis in this paper took place on an NVIDIA Titan XP gaming GPU with 3840 CUDA cores and 12GB GDDR5X RAM.

With our setup it is possible to run 3,000,000 iterations of the sampler with 14 months of data and 7 explanatory variables on a 256$\times$256 output grid, equivalent to 917,514 model parameters in around 6.5 hours. Our pilot work indicated that we could achieve an order of magnitude (25 times) speed up compared to our CPU implementation even on a lower specification NVIDIA GeForce GTX 1060 gaming GPU with 1280 CUDA cores and 6GB GDDR5/X. Thus using GPUs makes real-time outbreak monitoring for complex models, such as those described in the present article, possible - though note that the time needed in practice to run the algorithm will depend on mixing, which is data-dependent.

\section{Data Analysis}

We split our data into yearly chunks and analysed incidence at the monthly scale. Since we incorporated rainfall, EVI and temperature with a 1 month lag in our model, we needed to include at least one month of data from the previous year into our block, but we chose instead to include two months: thus each yearly block consisted of incidence and covariate data from (and including) November in the previous year up to December in the year of interest.

The population density data used for our offset were obtained from the WorldPop project \url{http://www.worldpop.org.uk}, aggregated onto our inferential grid. We included treatment-seeking rates into our model by multiplying the offset term, $P(s,t)$, by the treatment-seeking probabilities for each grid cell.

\subsection{Model Validation}

In order to test for goodness of fit, we compared the observed number of cases presenting to each facility to the predicted number of cases presenting to each facility as follows:
\begin{eqnarray*}
d &=& \frac{(\text{observed number of cases} - \text{predicted number of cases})}{\text{predicted number of cases}^{0.5}},\\
&=& \frac{O-E}{\sqrt E}
\end{eqnarray*}
We accounted for both health facility choice and treatment seeking behaviour in computing the predicted number of cases. The quantity $d$ in functional form resembles a signed version of the square root of a $\chi^2$ statistic.

Note that for validation purposes, we include the probability of seeking treatment in our model because we want to compare the number of cases predicted by our model to the actual number of observed cases. The latter is an underestimate of the real number of cases because of under-reporting. For prediction purposes it is better not to take into account treatment seeking, because this then yields an estimate of the true number of cases. In other words: for validation, we use as offset a product of the population and treatment seeking rasters, whereas for prediction, it is just the population raster; the latter is greater than or equal to the former.

\subsection{MCMC Details}

% nits <- 12000000
% burn <- 1000000
% thin <- 11000
% betaprior <- betapriorGauss(mean=0,sd=10000)
% omegaprior <- omegapriorGauss(mean=0,sd=1)
% etaprior <- etapriorGauss(mean=log(c(1,0.25)),sd=c(0.3,0.3))

We used zero mean independent Gaussian priors for $\beta$ with standard deviation $10^4$. For $\log\sigma$, $\log\phi$ and $\log\theta$, we used independent Gaussian priors with respective means 0, $\log{0.25}$ and 0 and respective standard deviations 0.3, 0.3 and 1. Since a distance of 1 degree in latitude at the equator equates to approximately 110km, our prior for $\phi$ gives a range of between 6--110km i.e. the ability to capture highly localised correlation and also long range correlation. The parameter $\sigma$ controls the variability in the latent spatial process, $\mathcal Y$, (which can be thought of as a sort of residual term) and the parameter $\phi$ controls the extent of the spatial correlation in $\mathcal Y$, larger values equate to greater correlation in space. Computation took place on a $256\times256$ inferential grid of size 0.047 degrees (approximately $5\times5$ km) covering the country. We ran the MCMC algorithm for 12,000,000 iterations, using a burnin of 1,000,000 iterations and retaining every 11,000th sample to give us a final sample of size 1000.

We checked for convergence and good mixing of our chain by examining the trace plot for each $\beta$ and $\eta$ parameter and we also used a plot of the log posterior density over the iterations as a global measure of convergence.

\section{Results}

Table \ref{tab:parameter-estimates} shows parameter estimates from each 14-month block of data together with a combined effect size calculated by merging posterior samples from each year.

With the exception of day and night-time temperature, all combined main effects were statistically significant at the 5\% level for the years 2013--2016. Considering individual years, however, our day-time temperature covariate showed a significant increasing relationship with incidence in 4 out of 5 years (2012,2014--2016) and a non-significant increasing relationship in 2013; indeed the overall effect size is only marginally insignificant - the lower end of the credible interval being 0.97. However, since we used a transform of day-time temperature, the interpretation of this regression coefficient is best appreciated by back-transformation -- see below. For night-time temperature, the relationship was significantly increasing in 2012 and 2013 and significantly decreasing in 2015 and 2016. Otherwise, our results show statistically significant increases in incidence with increasing rainfall, EVI and TWI.

\begin{table}[htbp]
    \centering
    \footnotesize
    \centering
    \begin{tabular}{|l|l|l|l}
        \textbf{Variable} & \textbf{2012} & \textbf{2013} & \textbf{2014} \\ \hline
        Intercept & 0.106* (3.82$\times10^{-2}$) & 0.114* (3.57$\times10^{-2}$) & 0.186* (5.26$\times10^{-2}$) \\
        Rainfall (m) & 1.5* (8.1$\times10^{-2}$) & 1.31* (5.6$\times10^{-2}$) & 1.28* (4.87$\times10^{-2}$) \\
        EVI & 1.11* (5.72$\times10^{-2}$) & 1.48* (6.06$\times10^{-2}$) & 1.57* (6.08$\times10^{-2}$) \\
        TempDay & 1.8* (0.186) & 1.08 (9.85$\times10^{-2}$) & 1.51* (0.128) \\
        TempNight (deg C) & 1.06* (1.33$\times10^{-2}$) & 1.03* (1.21$\times10^{-2}$) & 0.993 (1.08$\times10^{-2}$) \\
        TWI & 2.92* (0.293) & 2.88* (0.253) & 2.99* (0.273) \\
        sigma & 3.62 (5.37$\times10^{-2}$) & 3.64 (5.34$\times10^{-2}$) & 3.79 (5.36$\times10^{-2}$) \\
        phi & 7.38$\times10^{-2}$ (3.04$\times10^{-3}$) & 6.88$\times10^{-2}$ (2.7$\times10^{-3}$) & 5.96$\times10^{-2}$ (2.28$\times10^{-3}$) \\
        theta & 5.15 (0.188) & 5.12 (0.177) & 5.06 (0.166) \\
    \end{tabular}

    \begin{tabular}{|l|l|l|l}
        \textbf{Variable} & \textbf{2015} & \textbf{2016} & \textbf{Combined} \\ \hline
        Intercept & 0.176* (5.51$\times10^{-2}$) & 0.172* (5.68$\times10^{-2}$) & 0.151* (5.9$\times10^{-2}$) \\
        Rainfall (m) & 1.37* (6.92$\times10^{-2}$) & 1.4* (6.35$\times10^{-2}$) & 1.37* (1$\times10^{-1}$) \\
        EVI & 1.61* (6.99$\times10^{-2}$) & 1.78* (7.46$\times10^{-2}$) & 1.51* (0.232) \\
        TempDay & 1.22* (0.115) & 1.26* (0.107) & 1.38 (0.286) \\
        TempNight (deg C) & 0.971* (1.21$\times10^{-2}$) & 0.943* (1.06$\times10^{-2}$) & 1 (4.52$\times10^{-2}$) \\
        TWI & 3.21* (0.307) & 4.05* (0.393) & 3.21* (0.533) \\
        sigma & 3.59 (5.09$\times10^{-2}$) & 3.83 (5.99$\times10^{-2}$) & 3.69 (0.111) \\
        phi & 7.14$\times10^{-2}$ (2.69$\times10^{-3}$) & 5.3$\times10^{-2}$ (2.28$\times10^{-3}$) & 6.53$\times10^{-2}$ (8.25$\times10^{-3}$) \\
        theta & 5.09 (0.171) & 5.26 (0.227) & 5.14 (2$\times10^{-1}$) \\
    \end{tabular}
    \caption{\label{tab:parameter-estimates}Model parameter estimates (standard error) for each year. Combined estimates and credible intervals were obtained by merging posterior samples.}
\end{table}

\begin{figure}
    \centering
        \begin{minipage}{0.5\textwidth}
            \includegraphics[width=0.95\textwidth]{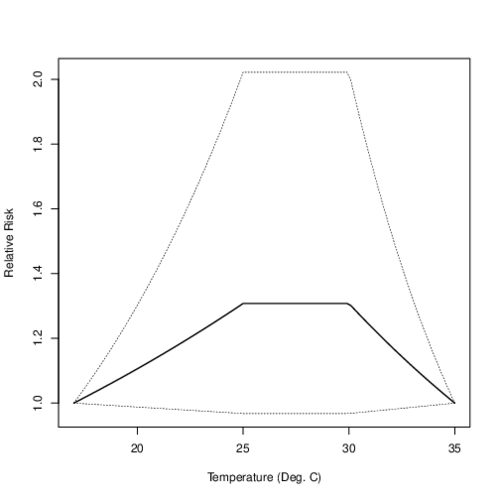}
        \end{minipage}
        % \begin{minipage}{0.5\textwidth}
        %     \includegraphics[width=0.95\textwidth]{fig/distrel}
        % \end{minipage}
    \caption{\label{fig:covrel} Illustrating the relationship between day-time temperature and relative risk.}
\end{figure}

The plot in Figure \ref{fig:covrel} shows the combined relationship between day-time temperature and risk. The relative risk, compared to baseline (in a nil transmission scenario i.e. below 17 degrees or above 35 degrees), in the `optimal' temperature zone (25--30 degrees) is 1.31 (95\%CRI 0.97-2.02); confidence bands for other temperature values can be read from the figure.

\begin{figure}
    \begin{center}
        \begin{minipage}{0.333\textwidth}
            \includegraphics[width=0.9\textwidth]{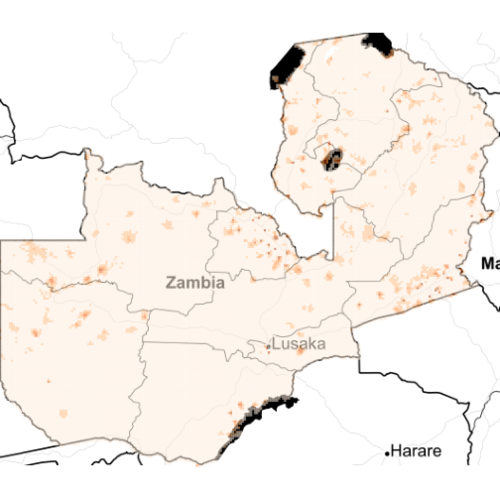}
        \end{minipage}\begin{minipage}{0.333\textwidth}
            \includegraphics[width=0.9\textwidth]{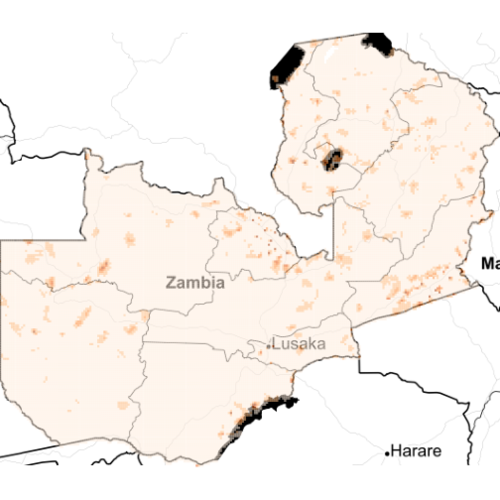}
        \end{minipage}\begin{minipage}{0.333\textwidth}
            \includegraphics[width=0.9\textwidth]{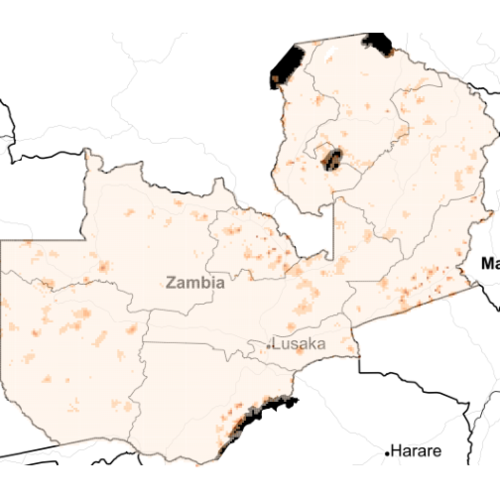}
        \end{minipage}

        \begin{minipage}{0.333\textwidth}
            \includegraphics[width=0.9\textwidth]{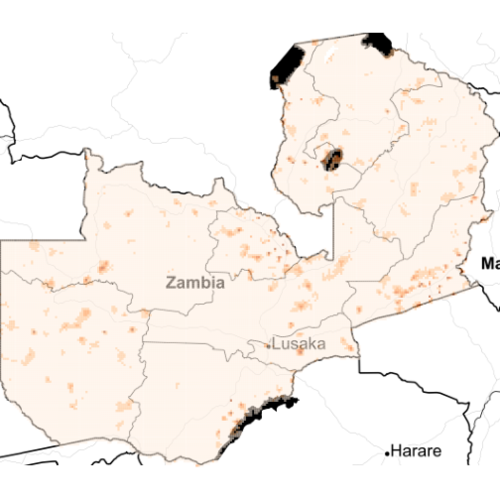}
        \end{minipage}\begin{minipage}{0.333\textwidth}
            \includegraphics[width=0.9\textwidth]{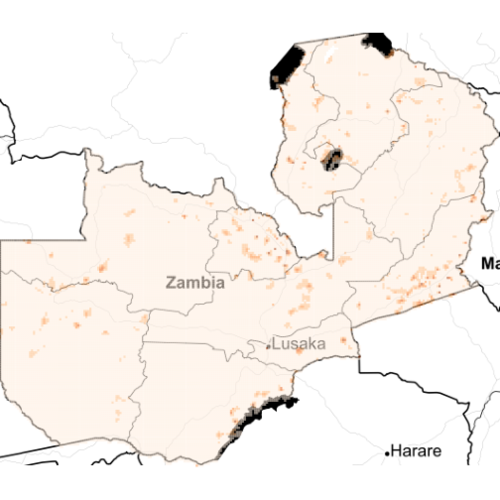}
        \end{minipage}\begin{minipage}{0.333\textwidth}
            \includegraphics[width=0.9\textwidth]{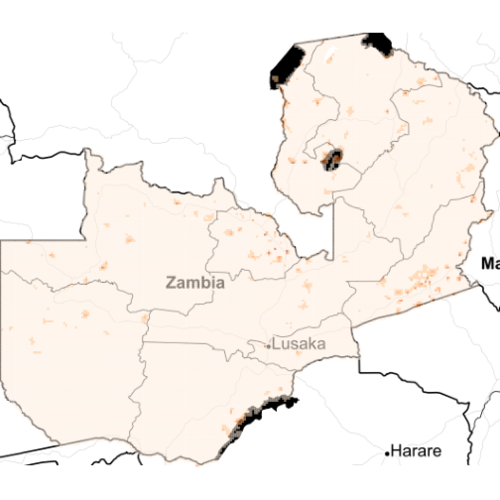}
        \end{minipage}

        \begin{minipage}{0.333\textwidth}
            \includegraphics[width=0.9\textwidth]{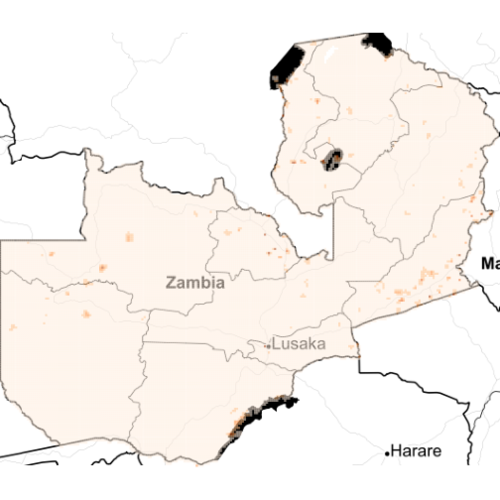}
        \end{minipage}\begin{minipage}{0.333\textwidth}
            \includegraphics[width=0.9\textwidth]{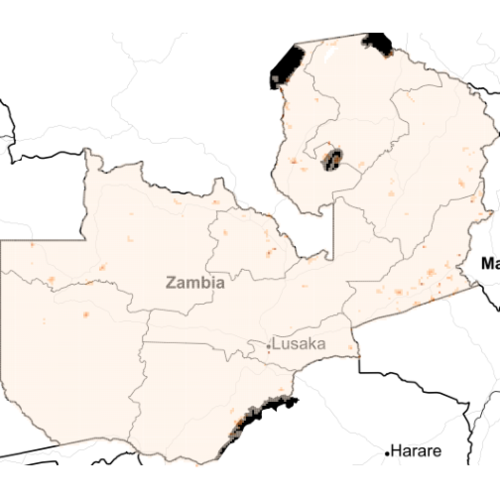}
        \end{minipage}\begin{minipage}{0.333\textwidth}
            \includegraphics[width=0.9\textwidth]{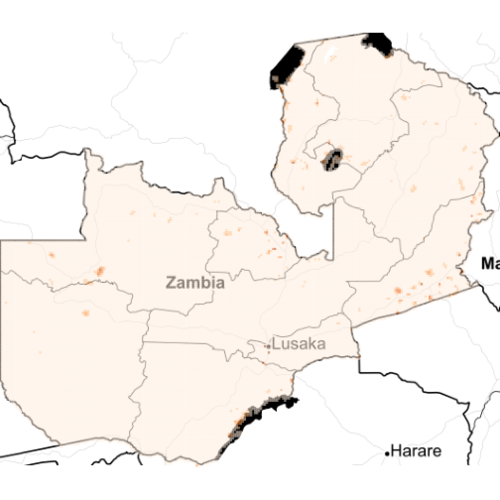}
        \end{minipage}

        \begin{minipage}{0.333\textwidth}
            \includegraphics[width=0.9\textwidth]{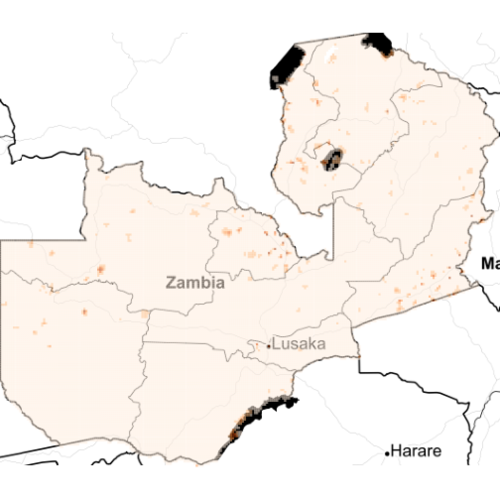}
        \end{minipage}\begin{minipage}{0.333\textwidth}
            \includegraphics[width=0.9\textwidth]{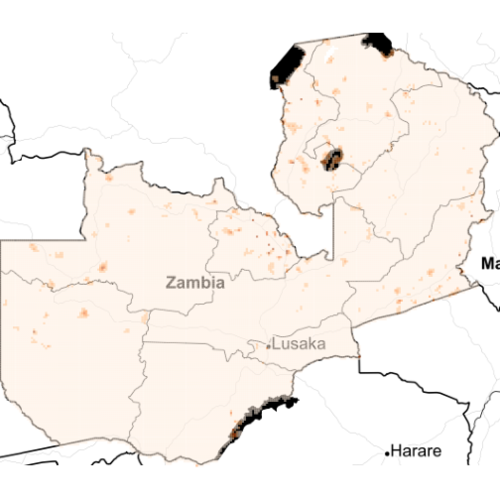}
        \end{minipage}\begin{minipage}{0.333\textwidth}
            \includegraphics[width=0.9\textwidth]{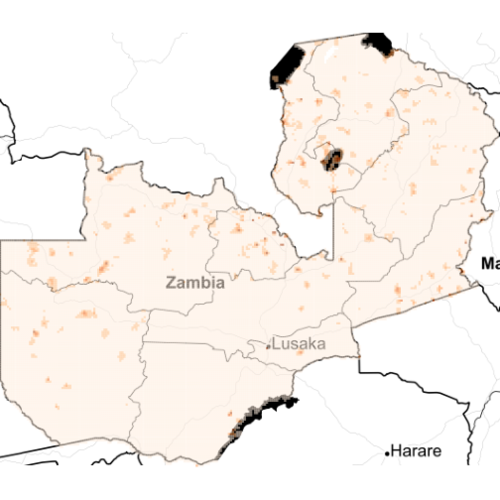}
        \end{minipage}

    \end{center}
    \caption{\label{fig:exceed_monthly} $\prob(\text{incidence $>$ 100 cases per 1000 people per month})$, left to right, top to bottom, the plots are for January -- December 2012. Colours are:
    \protect\includegraphics[width=1em]{col1} [ 0 , 0.2 ],
    \protect\includegraphics[width=1em]{col2} [ 0.2 , 0.4 ],
    \protect\includegraphics[width=1em]{col3} [ 0.4 , 0.6 ],
    \protect\includegraphics[width=1em]{col4} [ 0.6 , 0.8 ],
    \protect\includegraphics[width=1em]{col5} [ 0.8 , 1.0 ].
}
\end{figure}

\begin{figure}
    \begin{center}
        \begin{minipage}{0.4\textwidth}
            \includegraphics[width=0.9\textwidth]{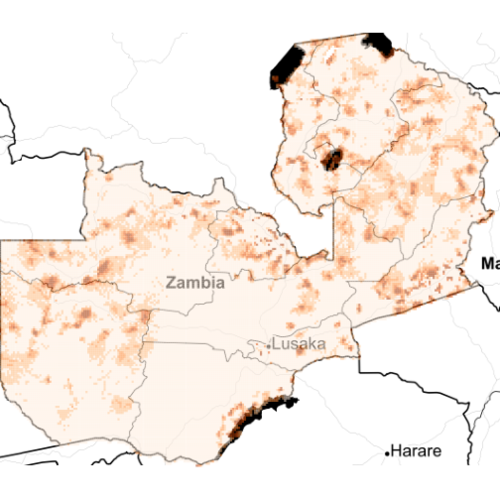}
        \end{minipage}\begin{minipage}{0.4\textwidth}
            \includegraphics[width=0.9\textwidth]{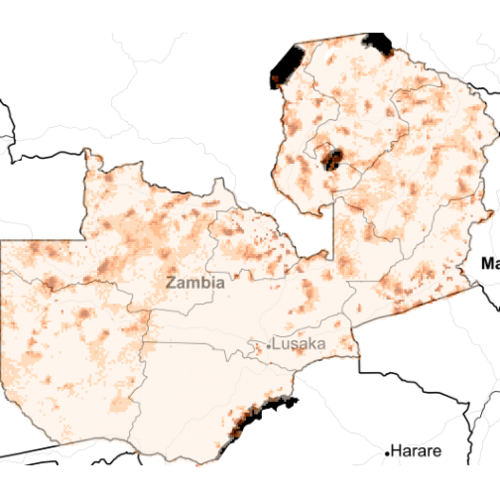}
        \end{minipage}

        \begin{minipage}{0.4\textwidth}
            \includegraphics[width=0.9\textwidth]{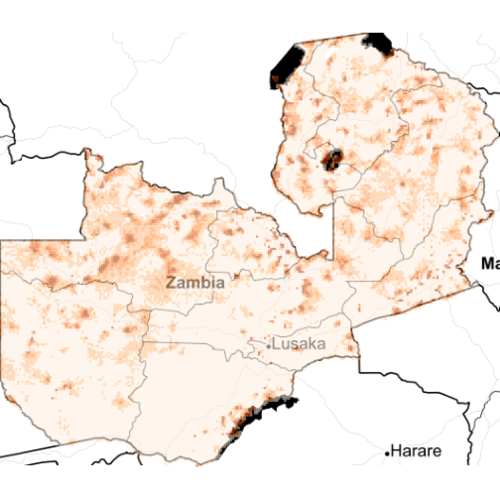}
        \end{minipage}\begin{minipage}{0.4\textwidth}
            \includegraphics[width=0.9\textwidth]{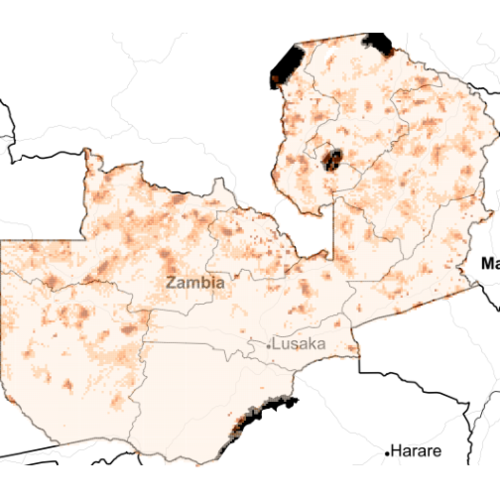}
        \end{minipage}

        \begin{minipage}{0.4\textwidth}
            \includegraphics[width=0.9\textwidth]{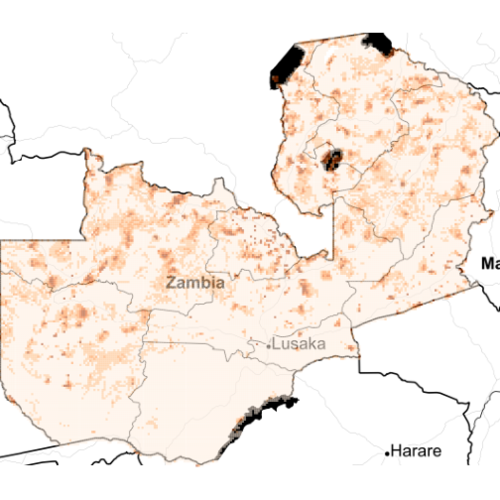}
        \end{minipage}

    \end{center}
    \caption{\label{fig:exceed_yearly}$\prob(\text{incidence $>$ 1200 cases per 1000 people per year})$ - note that $12\times100=1200$, left to right, top to bottom, the plots are for 2012 -- 2016. Colours are:
    \protect\includegraphics[width=1em]{col1} [ 0 , 0.2 ],
    \protect\includegraphics[width=1em]{col2} [ 0.2 , 0.4 ],
    \protect\includegraphics[width=1em]{col3} [ 0.4 , 0.6 ],
    \protect\includegraphics[width=1em]{col4} [ 0.6 , 0.8 ],
    \protect\includegraphics[width=1em]{col5} [ 0.8 , 1.0 ].}
\end{figure}

Figures \ref{fig:exceed_monthly} and \ref{fig:exceed_yearly} give
$$\prob(\text{incidence $>$ 100 cases per 1000 people per month})$$
for each month of the year in 2012 and
$$\prob(\text{incidence $>$ 1200 cases per 1000 people per year})$$
for each study year (note that $12\times100=1200$ i.e. both of these sets of plots are considering the same threshold). We present these exceedance probabilities, rather than the raw estimates of incidence because the exceedances take into account \emph{both} the estimated risk \emph{and} our ability to estimate it - we expect to be better able to estimate risk where there is a greater human presence, for instance. Darker areas in these plots are areas where we can be more confident that our estimate of risk exceeds the given thresholds.

Plots of $d$ versus the observed number of cases for each year are shown in Figure \ref{fig:validation} and indicate that we are very slightly overestimating the number of cases in areas where small numbers of cases are expected, and slightly underestimating the number of cases in areas where larger numbers of cases are expected.

In Figure \ref{fig:validation_spatial}, we have plotted the values of $d$ on a map (with $d$ averaged within the hexagonal markers). These plots show that our model has a slight tendency to overestimate the number of cases in the Eastern area of the country and in the North of the country, south of Lake Mweru; also that there is a slight tendency to underestimate the number of cases around Mumbwa, North West of Lusaka and in some areas of Southern Province.

\begin{figure}[htbp]
    \centering
        \includegraphics[width=0.5\textwidth]{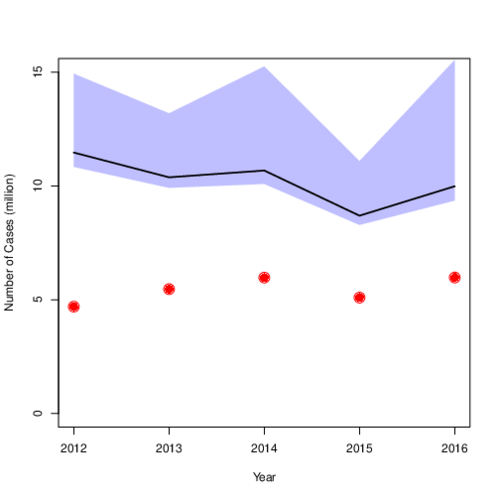}
    \caption{\label{fig:annual_incidence} Annual predicted number of cases of Malaria in Zambia. Line and credible interval show the predicted number of cases from our model, which accounts for underreporting. Red dots show the number of presumed and conformed cases in Zambia from the WHO World Malaria Report 2017 (pp 143).}
\end{figure}

Figure \ref{fig:annual_incidence} shows the predicted annual number of cases of malaria from 2012-2016, which in general shows a downward trend during the study period. The President's Malaria Initiative (\url{https://www.pmi.gov/}) estimates that 4.8 million cases of malaria were reported in 2016, so our modelling results also suggest that the true burden of malaria in Zambia is substantially underestimated based on health facility reports alone.

\section{Discussion}

Using aggregated health facility level data to model incidence and predict risk is challenging due to the uncertain catchment boundaries, rates of treatment seeking, and missing data. In this article, we present a rigorous solution to the problem of down-scaling malaria incidence derived from health facilities in Zambia in space and time that allows us to make predictions at fine-scale (community level) in a computationally efficient way. We modeled data from over 3,000 national reporting units over a five year period and found that community level incidence of malaria is likely substantially higher than that reported through the health system.

From a public health perspective, these results can be used to identify areas of high risk of disease across the entire country at a fine scale which may not be apparent in aggregate health system reports. Additionally, by presenting the results in terms of exceedance probabilities, our inference about disease incidence takes into account both the mean and uncertainty in incidence; thus we would only identify an area as high risk if (i) the estimated mean incidence was high relative to what would be predicted under the model and (ii) we were confident that our prediction was accurate. The latter will be the case in areas in which we are better able to estimate the incidence, i.e. in areas where health facilities are reporting on a consistent basis.

From a methodological perspective, the primary novelty and strength of our analysis is that each of the complicating factors described above is taken into account in a formal manner within our model, such that we are able to make correct inferences about the continuous disease process in space and time. A second strength is that we do not rely on fast, but approximate model fitting techniques: our method delivers inference from the full joint posterior of all model parameters - and with that, we can answer essentially any well-posed inferential question, see examples below.

This framework improves upon previous analyses that have included unsatisfactory assumptions, such as ignoring the fact that data pertain to areal units, or using unsatisfactory models to describe correlation structure between units, see \cite{wall2004} and \cite{taylor2015,taylor2018} for a detailed review and critique, including of the geostatistical technique of downscaling (e.g. area-to-point kriging). While there have been some recent efforts to address this, most are lacking in some way. For instance, \cite{nguyen2012} and \cite{lee2017}, use a local-EM approach for inference, and they consider both the changing boundaries problem and overlapping areal units in space and time, but their approach is based on a simple in-homogeneous Poisson process, rather than on doubly stochastic processes. \cite{wilson2018} consider combining point and areal data and \cite{alegana2016} and \cite{utazi2018} report methods for disaggregating respectively Poisson and binomial data with a joint modelling approach.

The main limitations on our study are broadly (1) those concerning the issue of missing data and (2) those concerning modelling assumptions. A third limitation is that we only used confirmed case data, though we did attempt to account for under-reporting in our analysis. Regarding (1): although our advocated modelling framework correctly accounts for missing values theoretically speaking, greater inherent completeness would certainly improve our ability to make predictions. But this is the reality of data collected by many nation health systems in developing country settings. Zambia has done a tremendous job of scaling up the collection and reporting of confirmed malaria case data over the past decade, but gaps remain especially in extremely remote areas. Regarding (2), our assumptions are split into those for (i) the treatment-seeking probability model, (ii) those for the catchment area model, and (iii) those for the data model. The main limitation for (i) is that our model is based on extrapolating spatial information from a smaller area to across the country. For (ii) and (iii), we respectively assume a Multiplicative Competitive Interaction model and a log-Gaussian Cox process; while it is at best difficult to assess the validity of the MCI model, we have endeavoured to assess the fit of the LGCP. Regarding the latter, one feature not explicitly captured by our model is the potential of malaria cases to cause other local cases via vector transmission; however, the incorporation of a spatially-correlated random effects term attempts to capture these local outbreaks. Future research in this area could also consider in greater detail the form of the relationship with putative covariates.

These limitations aside, the framework and the outputs presented will provide the Zambia National Malaria Elimination Centre and implementing partners increased ability to target interventions spatially at levels below the health catchment area, and to integrate these models with country surveillance data for improved data-driven decision making. In addition, we have not illustrated in this article the full power of our inferential framework for Zambia and elsewhere. This is that given an arbitrary set of aggregation units our method can compute unbiased estimates of the mean, variance, maximum, minimum or indeed any function of our model parameters (including incidence or relative risk) on these units. Examples of programmatically relevant questions that can be couched in these terms include 1) predicting the number of cases likely to report to a new health facility location; 2) estimating the predicted effect (with confidence intervals) of different intervention distribution plans on incidence; and 3) including information on cost to assess the cost-effectiveness of different intervention mixes.

\section*{Acknowledgements}

The authors are grateful to Luigi Sedda and Emanuele Giorgi for helpful comments during the preparation of this manuscript.

\section*{Ethics, Consent and Permissions}

This study used retrospective secondary data pertaining to humans, but none of the data were personally identifiable - we only used counts of malaria at the health facility level. Ethics for this study was obtained by Dr John Miller and Dr Thomas Eisele, from The University of Zambia (refs 007-03-14 and 026-09-18) and Tulane University (ref 14-592085).

\section*{Competing Interests}

In preparing this manuscripot, none of the authors have any competing interests to declare.

\bibliographystyle{plain}
\bibliography{zambiamalariaarxiv}

\appendix

\section{Facility Choice Model and Data Augmentation \label{sect:huff_data}}

Multiplicative Competitive Interaction models are built around the concept of utility. For the purpose of this work, utility means the attraction of a patient towards one of the possible health facilities. These utilities are defined as non-normalized probabilities, so
that if $u_{ij}$ is the utility that individual $i$ gets from attending health facility $j$, then the individual's choice probability is given by
\begin{equation*}
  q_{ij} = \frac{u_{ij}}{\sum_j u_{ij}}.
  \label{eq:hf_preference}
\end{equation*}

MCI models assume that the utility can be expressed as a function of the competing choice's attributes.
We use attributes based on location and type of the health facilities,
and express the utility as
\begin{equation*}
  u_{ij} = e^{\tau_j} \prod_{k=1}^2 B^{\beta_k}_{k(ij)},
  \label{eq:utility}
\end{equation*}
where $\tau_j$ is a parameter that depends on the type of health
facility\footnote{In Zambia health facilities are classified into hospitals,
health centers, health posts and community health workers. The characteristics
of these categories depend on the size, location and medical services they
provide.}, $B_{1(ij)}$ is the travel time required by patient $i$ to visit
facility $j$, $B_{2(ij)}$ is the number of health facilities that require a lower travel cost than facility $j$, and $\beta_k$ are calibration parameters.

We can use the geometric mean ($\tilde{\mu}$) as normalization constant for the preferences
and express the probability of attendance as a linear function of the health
facility attributes, as follows
\begin{equation}
   \log \left( \frac{q_{ij}}{\tilde{\mu}(q_{i:})} \right) = \tau_j +
           \sum_{k=1}^2 \beta_k \log \left(\frac{B_{k(ij)}}{\tilde{\mu}(B_{k(i:)})} \right).
  \label{eq:mci}
\end{equation}

To calibrate a model like Eq~\ref{eq:mci} we would need information about the health facilities attributes, the travel time from the patients household to the health facilities and the patients preferences ($q_{ij}$). Ideally patients preferences would be expressed as a frequency of visits to one or another health facility. However, the survey data used does not contain such information, but only one visit per patient to a single facility.
We developed an heuristic mechanism to generate the data needed and and calibrate the MCI model. This mechanism was implemented using only on 80\% of the data, while the remaining of 20\% was used to assess that the resulting predicted preferences  were  aligned  to  the  survey  data.

We assumed that the frequencies and pattern of visits observed across all patients are representative of each one of them. In other words, we assumed that all patients follow the same decision process.
Table~\ref{tab:visit_frequencies} shows the frequencies of visits to each health facility per type and travel time required to get to them.
The time intervals in the Table were chosen according to the deciles of the travel times to the facility visited by each patient. The frequencies in the Table were scaled considering all facilities available to each patient, and not only the one visited.
Under our assumption, all patients visit different health facilities according to the frequencies shown. However, not all patients have a access to health facility of each type and time travel interval from Table \ref{tab:visit_frequencies}. In such cases the frequencies were re-scaled to match the actual health facilities available.
Table~\ref{tab:preference_algo} explains each step followed to generate a set of pseudo-observations of the frequencies per patient.

\begin{table}[ht]
  \centering
  {\small
  \begin{tabular}{c|ccccc}
    \hline
      & 0 - 7 & 7 - 18 & 18 - 27 & 27 - 43 & 43 - 66 \\
    \hline
    C & 20.01\%   & 10.57\%       & 10.98\%       & 4.84\%        & 3.76\%       \\
    P & --        &  6.87\%       &  9.16\%       & 8.02\%        & 9.16\%       \\
    \hline
    \hline

      & 66 - 113 & 113 - 159 & 159 - 280 & 280 - 508 & 508 - 1505 \\
    \hline
    C &  1.93\%      & 1.55\%      & 0.33\%      & 0.11\%     & 0.06\% \\
    P &  4.58\%      & 5.44\%      & 2.37\%      & 0.25\%     & $<$0.01\% \\
    \hline

  \end{tabular}
  }
  \caption{Frequency of visits to health facilities per travel time bracket (measured in minutes) and
  type of facility: health centers (C) and health posts (P).}
  \label{tab:visit_frequencies}
\end{table}

\begin{table}[ht]
  \centering
  \begin{tabular}{p{.1cm} p{11cm}}
    \hline
    \multicolumn{2}{l}{Steps followed:}  \\
    \hline
    \texttt{1} & Define a list $\mathcal{L}$ of all health facilities in the survey  \\
    \texttt{2} & For each patient in the survey: \\
    \texttt{3} & \hspace{.5cm} Compute travel times to $\mathcal{L}$ from patient's location \\
    \texttt{4} & \hspace{.5cm} Allocate facilities in $\mathcal{L}$ to quadrants in Table~\ref{tab:visit_frequencies} \\
    \texttt{5} & \hspace{.5cm} Assign the frequencies of Table Table~\ref{tab:visit_frequencies} to facilities in $\mathcal{L}$\\
    \texttt{6} & \hspace{.5cm} If multiple facilities are in the same quadrant:\\
    \texttt{7} & \hspace{1.0cm} Split the frequencies evenly\\
    \texttt{8} & \hspace{.5cm} Re-scale the frequencies so that they add up to one\\
    \hline
  \end{tabular}
  \caption{Proceedure to generate frequencies.}
  \label{tab:preference_algo}
\end{table}

We used the pseudo-observations to calibrate the Huff model described in Eq. \ref{eq:mci} and parameterize individuals preferences. Table \ref{tab:huff_regression} shows the
parameter estimates.

\begin{table}[htbp]
    \footnotesize
    \centering

    \begin{tabular}{|l|r|r}
        Variable               & Estimate  & p-value\\
        \hline
        $\tau$ - Health Center & -1.02     & $<$2e-16 \\
        $\tau$ - Health Post   &  0.81     & $<$2e-16 \\
        $B_1$ - Travel Time        & -1.99     & $<$2e-16 \\
        $B_2$ - Num of Closer Facilities  & -0.60     & $<$2e-16 \\
    \end{tabular}
    \caption{Model parameter estimates for health facility preferences.  \label{tab:huff_regression}}
\end{table}

The preference estimates $q_{ij}$ do not depend only on the parameters in Table \ref{tab:huff_regression}. The linear model  in Eq. \ref{eq:mci} is applied on $q_{ij}$ scaled by the harmonic mean $\tilde{\mu}(q_{i:})$. Then the preferences will depend on the options available. This makes the interpretation of the parameters in Table \ref{tab:huff_regression} somehow involved. However we can illustrate how the model works with some examples. The parameters $\tau$ reflect a preference of attending Health Posts over Health Centers. A patient with access to a health center and a health post with the same characteristics, so that both incur in the same travel time and have the same number of closer competitors, will attend the health post $70\%$ of the cases and the health center only $30\%$.

\begin{figure}
    \centering
        \begin{minipage}{0.4\textwidth}
            \includegraphics[width=0.9\textwidth]{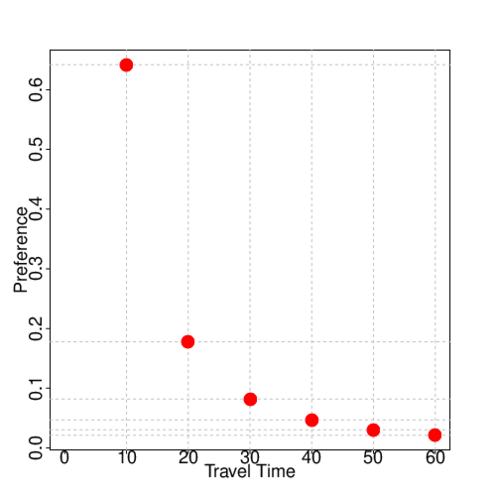}
        \end{minipage}\begin{minipage}{0.4\textwidth}
            \includegraphics[width=0.9\textwidth]{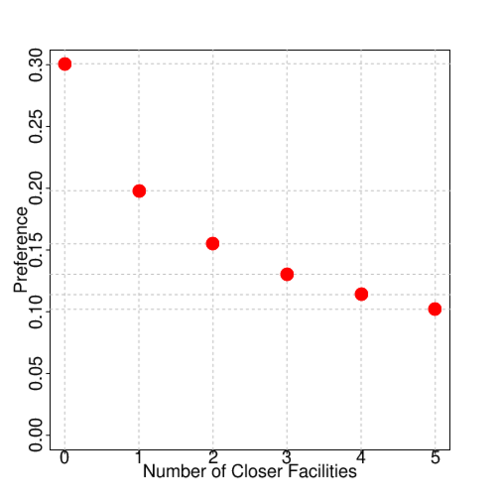}
        \end{minipage}
    \caption{\label{fig:huff_decay}Patients preferences towards 6 health facilities spaced every 10 minutes.
    Left, assuming no effect of closer competitors. Right, assuming no effect of travel time.}
\end{figure}

Consider a patient who has access to 6 health facilities of the same type, each one every 10 minutes away. Without considering the effect of the number of closer competing health facilities (i.e. assuming $\beta_2 = 0$), the preference of the patient towards each facility decays as shown in Fig. \ref{fig:huff_decay} on the left side. Conversely, now assume that the effect of travel time is nil (i.e. $\beta_1 = 0$) and that the only driver of the preferences is the number of closer competing facilities. The preference of the patient towards the 6 facilities decreases as the number of closer competitors increase according to Fig. \ref{fig:huff_decay} on the right side.

We predicted the health facility most likely to be visited by each patient in the 20\% of data that was initially kept aside. Depending on the household location, each individual in the validation set could choose from a varied set of alternatives, ranging from 50 to over 4000, with 162 being the mean.
We found that in 56.19\% of the cases the facilities visited matched the ones with the highest preference predicted by our model. Considering the number of health facility alternatives per per patient, this result suggest that the estimated preferences are consistent with the survey data. While this is far from being a formal validation, this is as far as we can go at this stage, given the lack of data to carry on other type of tests. This model allows us to define patients preferences in a multinomial setting, following an intuitive approach, which can then be incorporated as an initial assumption in a Bayesian framework.

\begin{figure}
    \centering
        \begin{minipage}{0.4\textwidth}
            \includegraphics[width=0.9\textwidth]{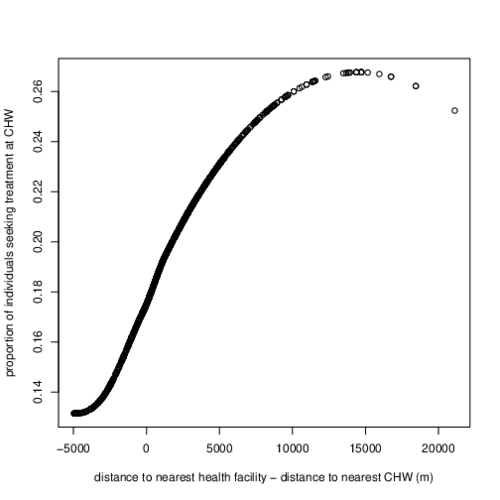}
        \end{minipage}
    \caption{\label{fig:tseek_chw} Among all individuals who sought care for fever in the past two weeks, proportion of individuals who sought care at a CHW, plotted against the difference in distance between the nearest health facility and the nearest active CHW.}
\end{figure}

Figure \ref{fig:tseek_chw} shows the increase in treatment seeking at a CHW, among those who sought treatment, when distance to the nearest health facility is relatively further away than the nearest active CHW.

\section{Handling of Missing Values \label{sect:miss_val}}

In this section, we explain in detail how missing values were handled in our analysis.

\begin{figure}
    \centering
    \includegraphics[width=0.7\textwidth]{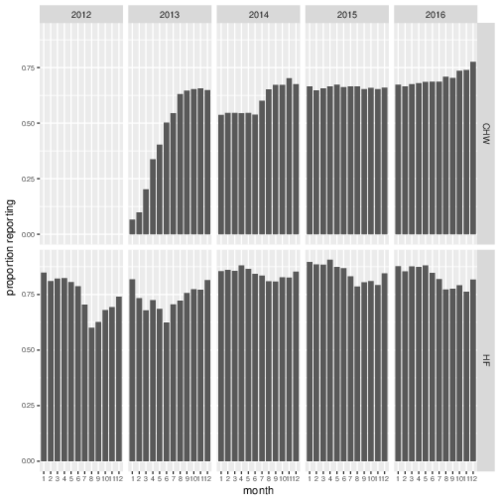}
    \caption{\label{fig:reporting} Bar chart showing the proportion of community health workers (CHWs) and health facilities (HF) reporting
    by month over the study period.}
\end{figure}

Figure \ref{fig:reporting} shows the proportion of community health workers and health facilities reporting per reporting month.

\begin{figure}[H]
    \centering
    \begin{minipage}{0.45\textwidth}
        \includegraphics[width=\textwidth]{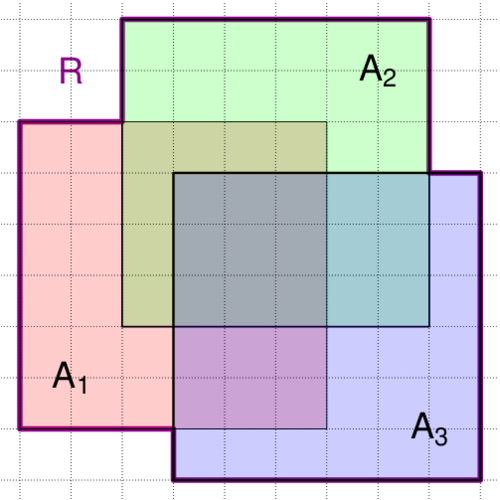}
    \end{minipage}\hfill\begin{minipage}{0.45\textwidth}
        \includegraphics[width=\textwidth]{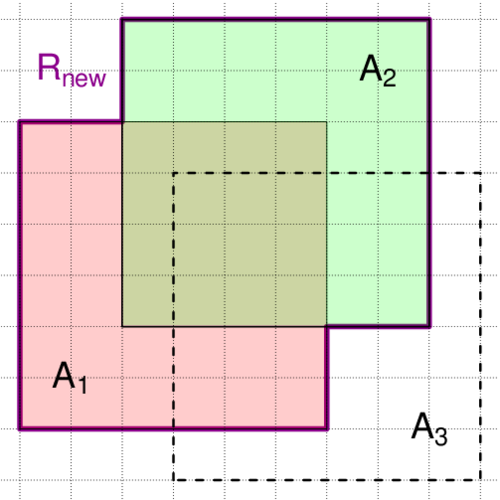}
    \end{minipage}
    \caption{\label{fig:NAfig}Figure illustrating the method for handling missing values. Dotted lines represent the grid cells in our analysis -- see text for further detail. In the left diagram, we have complete data from each of the figures and in the right diagram, one of the facilities, $A_3$ did not report. The observation window in each case is bounded by a bold purple line. If disease counts at facility $A_3$ are not observed then in our analysis, we assume that individuals that would have reported to $A_3$ would continue to do so. Therefore we shrink the observation window and seek to reduce the proportion of reporting people in the region $A_3\cap(A_1\cup A_2)$. If we did not reduce the proportion of reporting people then in the right hand diagram, then there would seem to be an under-reporting of cases in facilities $A_1$ and $A_2$ because there would be more people reporting than there really are (those individuals in fact reported to $A_3$, but we did not observed this).}
\end{figure}

Figure \ref{fig:NAfig} illustrates our method for handling missing values. In this figure, coloured regions are health facility catchments and dotted lines are the computational grid, on which we have information about sub-catchment level population. In the left plot, the observation window $R$ is partitioned into three overlapping health facility catchment areas $A_1$, $A_2$ and $A_3$. Suppose the expected number of cases arising any grid cell $i$ is $E_i$; in a typical Poisson model, this would be proportional to the number of people at risk of the disease and would act, as it does in our case, as an `offset'.

We proceed by:
\begin{enumerate}
    \item Shrinking the observation window from $R$ in the left plot to $R_{\text{new}}$ in the right plot.
    \item Shrinking $E_i$ for each cell $i$ still in the intersection $A_3\cap(A_1\cup A_2)$ and setting $E_i=0$ for cells outside it as we no longer observe the disease process there.
\end{enumerate}
Note that if we did not shrink $E_i$ in the intersection $A_3\cap(A_1\cup A_2)$, then in the right hand case, there would be `too-many' people in the observation window, thus we need to reduce this number in a formal way to account for the fact that these individuals have not disappeared or changed their reporting behaviour, but continue to report to $A_3$.

Formally, let $w_{ij}$ be the $[i,j]$ element of the matrix $W$ and recall that by definition,
\begin{equation} \label{eqn:condprob}
    w_{ij} = \P\{\text{an individual attends facility $j$} | \text{they are in cell $i$}\}
\end{equation}
Since for cells $i$ that report to \emph{some} health facility, we have $\sum_j w_{ij} = 1$, it is straightforward to see that we should adjust $E_i$ to
\begin{equation*}
    E_i^\star = \left(1-\sum_{j\in\mathcal{N}} w_{i,j}\right)E_i,
\end{equation*}
where $\mathcal{N}$ is the set of facilities that did not report (but should have) at each relevant time point.

\section{Additional Plots}

\begin{figure}
    \begin{center}
        \begin{minipage}{0.4\textwidth}
            \includegraphics[width=0.9\textwidth]{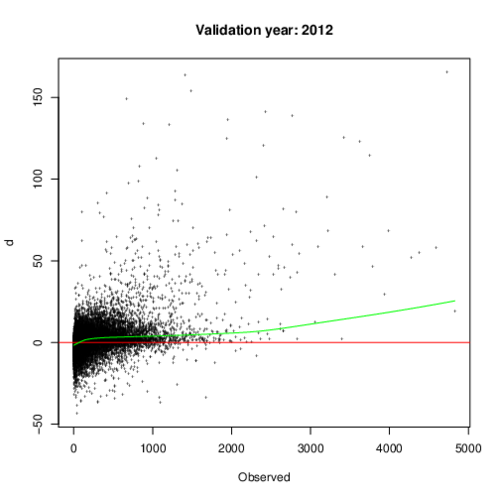}
        \end{minipage}\begin{minipage}{0.4\textwidth}
            \includegraphics[width=0.9\textwidth]{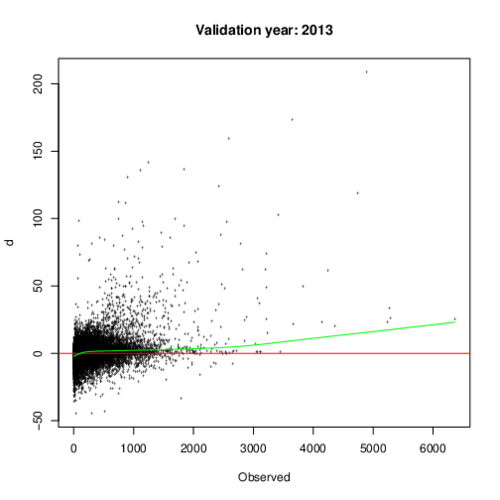}
        \end{minipage}

        \begin{minipage}{0.4\textwidth}
            \includegraphics[width=0.9\textwidth]{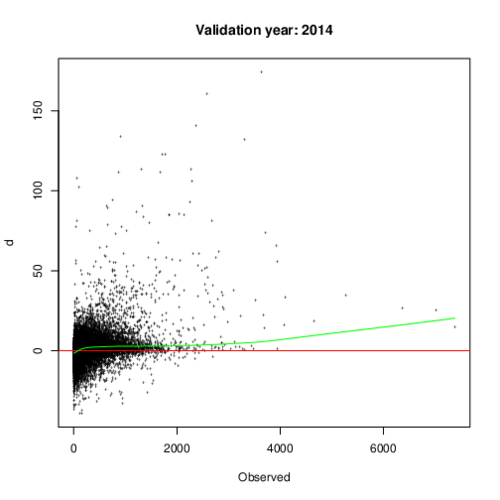}
        \end{minipage}\begin{minipage}{0.4\textwidth}
            \includegraphics[width=0.9\textwidth]{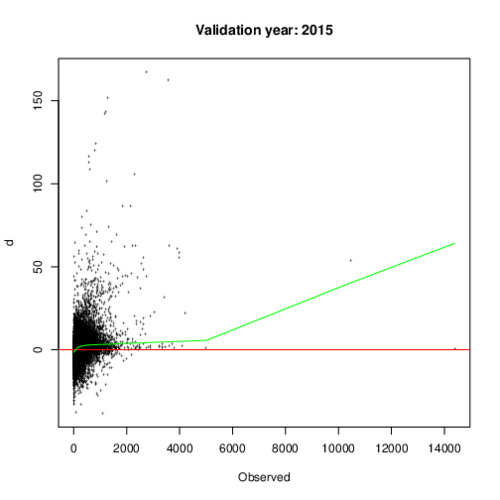}
        \end{minipage}

        \begin{minipage}{0.4\textwidth}
            \includegraphics[width=0.9\textwidth]{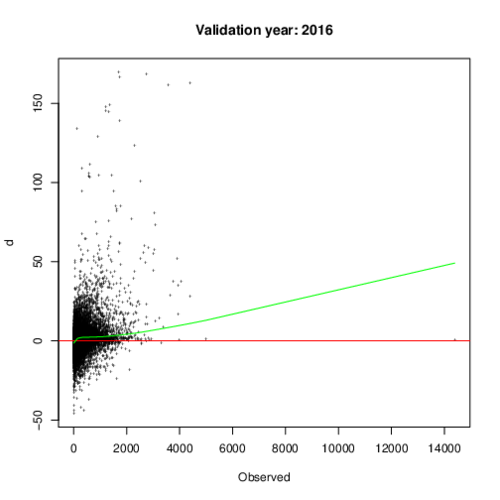}
        \end{minipage}
    \end{center}
    \caption{\label{fig:validation}Validation plots. For each facility, we calculate $d = (\text{observed number of cases} - \text{predicted number of cases}) / \text{predicted number of cases}^{0.5}$. Red line is $d=0$ and green line is a lowess smooth. Left to right, top to bottom, the plots are for 2012 -- 2016. See text for further detail.}
\end{figure}

\begin{figure}
    \begin{center}
        \begin{minipage}{0.4\textwidth}
            \includegraphics[width=0.9\textwidth]{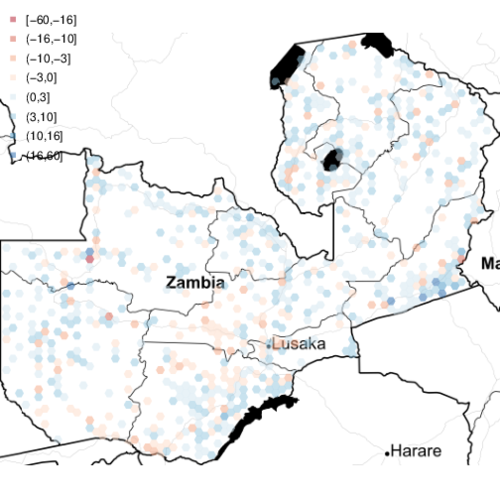}
        \end{minipage}\begin{minipage}{0.4\textwidth}
            \includegraphics[width=0.9\textwidth]{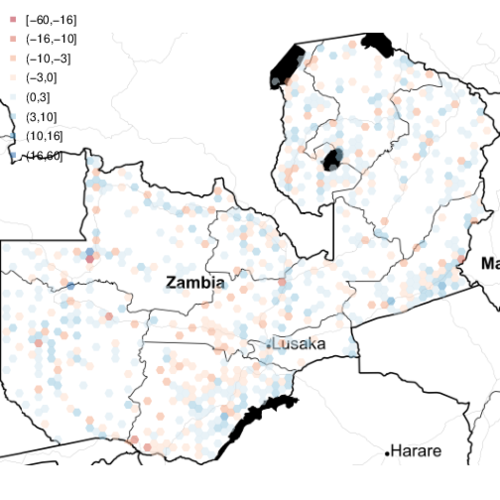}
        \end{minipage}

        \begin{minipage}{0.4\textwidth}
            \includegraphics[width=0.9\textwidth]{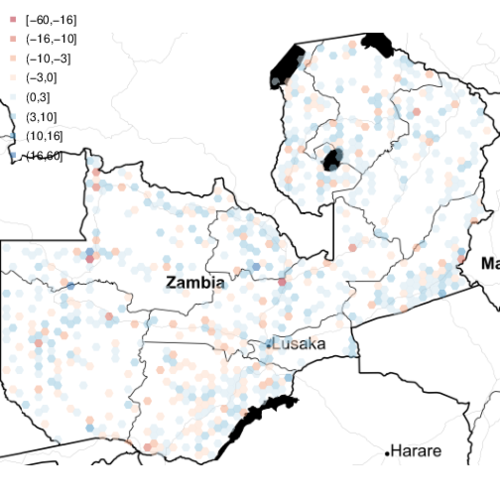}
        \end{minipage}\begin{minipage}{0.4\textwidth}
            \includegraphics[width=0.9\textwidth]{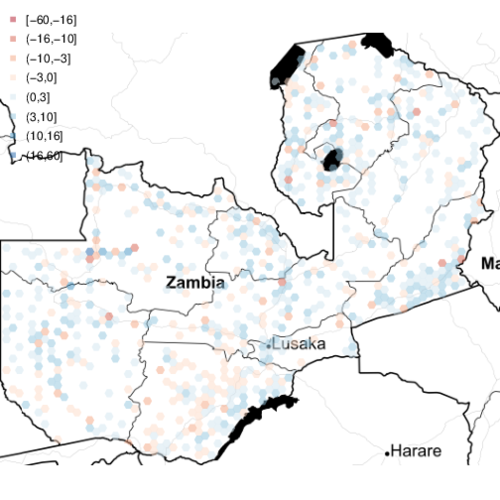}
        \end{minipage}

        \begin{minipage}{0.4\textwidth}
            \includegraphics[width=0.9\textwidth]{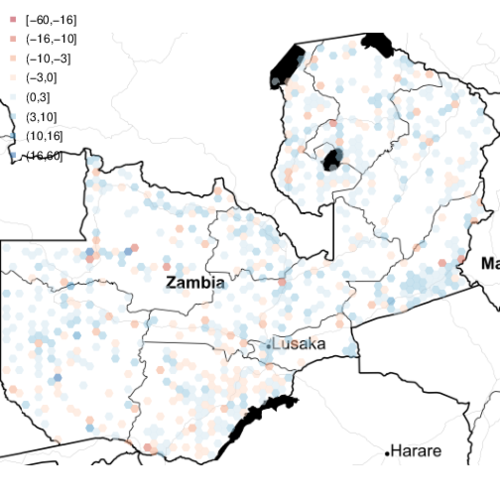}
        \end{minipage}
    \end{center}
    \caption{\label{fig:validation_spatial}Spatial validation plots, showing where incidence was over- and under-estimated. For each facility, we calculate $d = (\text{observed number of cases} - \text{predicted number of cases}) / \text{predicted number of cases}^{0.5}$. Left to right, top to bottom, the plots are for 2012 -- 2016. See text for further detail.}
\end{figure}

\end{document}